\documentclass[12pt]{article}
\usepackage{graphics}
\begin{document}

\title{Quantum probabilistic teleportation via entangled coherent states}
\author{Shang-Bin Li and Jing-Bo Xu\\
Chinese Center of Advanced Science and Technology (World Labor-\\
atory), P.O.Box 8730, Beijing, People's Republic of China;\\
Zhejiang Institute of Modern Physics and Department of Physics,\\
Zhejiang University, Hangzhou 310027, People's Republic of
China\thanks{Mailing address}}
\date{}

\maketitle

\begin{abstract}
{\normalsize We study the entanglement of entangled coherent
states in vacuum environment by employing the entanglement of
formation and propose a scheme to probabilistically teleport a
coherent superposition state via entangled coherent states, in
which the amount of classical information sent by Alice is
restricted to one bit. The influence of decoherence due to photon
absorpation is considered. It is shown that decoherence can
improve the mean fidelity of probabilistic teleportation in some situations.\\

}

\end{abstract}
\newpage
\section * {I. INTRODUCTION}
\hspace*{8mm}Quantum entanglement plays an important role in
various fields of quantum information, such as quantum
computation[1], quantum teleportation [2,3], dense coding [4] and
quantum communication [5], etc. Entanglement can exhibit the
nature of a nonlocal correlation between quantum systems that have
no classical interpretation. It has been noted [2,3] that quantum
teleportation can be viewed as an achievable experimental
technique to quantitatively investigate quantum entanglement.
Several quantum teleportation schemes of both discrete and
continuous variables have been proposed [2,3,6,7]. Continuous
variable quantum teleportation of an unknown coherent states has
been realized experimentally by employing a two-mode squeezed
vacuum state as an entanglement resource [3]. Recently, the
teleportation schemes via the entangled coherent
states have been studied [8].\\
\hspace*{8mm}In this Letter, we study the entanglement of
entangled coherent states in vacuum environment by employing the
entanglement of formation [9] and find that the entanglement of
formation of the entangled coherent states is sensitive with the
relative phase $\phi$ when the amplitude $|\alpha|$ is very small.
Then, we propose a scheme of probabilistic teleportation via
entangled coherent states, in which the amount of classical
information sent by Alice is restricted to one bit. In this
probabilistic teleportation scheme, a coherent superposition state
can be probabilistically perfectly teleported via a properly
chosen entangled coherent state. When the interaction with the
vacuum environment is addressed, the mean fidelity of the scheme
is studied. It is shown that the decoherence due to interaction
with the environment can improve the mean fidelity in some
specific situations.\\

\section * {II. ENTANGLEMENT OF ENTANGLED COHERENT STATES}
\hspace*{8mm}In this section, we investigate the entanglement
properties of entangled coherent states by analyzing the
entanglement of formation. The pure entangled coherent states is
defined by,
$$
|C(\alpha_1,\alpha_2,\phi)\rangle=N_{\phi}(|\alpha_1\rangle_1|\alpha_2\rangle_2
+e^{i\phi}|-\alpha_1\rangle_1|-\alpha_2\rangle_2),
\eqno{(1)}
$$
where the normalization constant
$N_{\phi}=[2+2\cos\phi\exp(-2|\alpha_1|^2-2|\alpha_2|^2)]^{-1/2}$
and $|\pm\alpha_{1,2}\rangle$ are coherent states. \\
\hspace*{8mm}Many measures of entanglement have been introduced
and analyzed [9,10,11]. Here, we adopt the entanglement of
formation to analyze the entanglement properties of pure or mixed
entangled coherent states. The entanglement of formation of pure
entangled coherent states $|C(\alpha_1,\alpha_2,\phi)\rangle$
reduces to the von Neumann entropy of the reduced density matrix
$\rho_2\equiv\textrm{Tr}_1(|C(\alpha_1,\alpha_2,\phi)\rangle\langle{C}(\alpha_1,\alpha_2,\phi)|)$,
i.e.,
$E(|C(\alpha_1,\alpha_2,\phi)\rangle)=-\textrm{Tr}\rho_2\log_2\rho_2$.
The reduced density matrix $\rho_2$ can be expressed as
$$
\rho_2\equiv\textrm{Tr}_1(|C(\alpha_1,\alpha_2,\phi)\rangle\langle{C}(\alpha_1,\alpha_2,\phi)|)
$$
$$
~~~=N^2_{\phi}[1+\exp(-2|\alpha_1|^2)][1+\cos\phi\exp(-2|\alpha_2|^2)]
|\chi_{+}\rangle\langle\chi_{+}|
$$
$$
~~~~+N^2_{\phi}[1-\exp(-2|\alpha_1|^2)][1-\cos\phi\exp(-2|\alpha_2|^2)]|\chi_{-}\rangle\langle\chi_{-}|,
\eqno{(2)}
$$
where
$$
|\chi_{+}\rangle=[2+2\cos\phi\exp(-2|\alpha_2|^2)]^{-1/2}(|\alpha_2\rangle+e^{i\phi}|-\alpha_2\rangle),
$$
$$
|\chi_{-}\rangle=[2-2\cos\phi\exp(-2|\alpha_2|^2)]^{-1/2}(|\alpha_2\rangle-e^{i\phi}|-\alpha_2\rangle).
\eqno{(3)}
$$
It is easy to obtain the two non-negative eigenvalues
$\lambda_{\pm}$ of $\rho_2$ as follows,
$$
\lambda_{\pm}=\frac{1}{2}\pm\sqrt{\frac{1}{4}-N^4_{\phi}[1-\exp(-4|\alpha_1|^2)][1-\exp[(-4|\alpha_2|^2)]}.
\eqno{(4)}
$$
Then, the entanglement of formation of entangled coherent states
$|C(\alpha_1,\alpha_2,\phi)\rangle$ is given by
$$
E(|C(\alpha_1,\alpha_2,\phi)\rangle)=-\lambda_{+}\log_2\lambda_{+}-\lambda_{-}\log_2\lambda_{-}.
\eqno{(5)}
$$
The entanglement of formation $E$ of the entangled coherent states
with the relative phase $\phi=\pi$ is shown in Fig.1. In Fig.2, we
plot the entanglement of formation of the entangled coherent
states $|C(\alpha_1,\alpha_2,\phi)\rangle$ as the function of
$|\alpha_1|=|\alpha_2|=|\alpha|$ and the relative phase $\phi$. It
is shown that $|C(\alpha,\alpha,\pi)\rangle$ is a maximally
entangled pure state in $2\times 2$ Hilbert space. When, the
relative phase $\phi$ varies from $0$ to $2\pi$, the degree of
entanglement firstly increases, and achieve a maximal value,
then decreases to a minimal value. \\
\hspace*{8mm}Next, we discuss the influence of decoherence on the
entanglement of $|C(\alpha_1,\alpha_2,\phi)\rangle$. When the
entangled coherent states $|C(\alpha_1,\alpha_2,\phi)\rangle$ are
disposed to a vacuum environment, the states decohere and become
mixed state $\rho(t)$, where $t$ stands for the decoherence time.
The time evolution of $\rho(t)$ satisfies the following master
equation [12],
$$
\frac{\partial\rho(t)}{\partial{t}}=\sum^{2}_{i=1}\gamma\hat{a}_i\rho(t)\hat{a}^{\dagger}_i-
\frac{\gamma}{2}(\hat{a}^{\dagger}_i\hat{a}_i\rho(t)+\rho(t)\hat{a}^{\dagger}_i\hat{a}_i),
\eqno{(6)}
$$
where $\gamma$ is the decay rate, and $\hat{a}_{i}$,
$\hat{a}^{\dagger}_{i}$ ($i=1,2$) are the $i$th mode annihilation
and creation operators. The solution of the master equation can be
obtained
$$
\rho(t)=\sum^{\infty}_{n_1,n_2=0}\frac{(1-e^{-\gamma{t}})^{n_1+n_2}}{n_1!n_2!}\hat{L}(t)
\hat{a}^{n_1}_1\hat{a}^{n_2}_2\rho(0)\hat{a}^{\dagger{n_1}}_1\hat{a}^{\dagger{n_2}}_2\hat{L}(t),
\eqno{(7)}
$$
where
$\hat{L}(t)=\exp[-\frac{\gamma}{2}(\hat{a}^{\dagger}_1\hat{a}_1+\hat{a}^{\dagger}_2\hat{a}_2)t]$.
Substituting
$\rho(0)=|C(\alpha_1,\alpha_2,\phi)\rangle\langle{C}(\alpha_1,\alpha_2,\phi)|$
into the Eq.(7), we can express the mixed state $\rho(t)$ as,
$$
\rho(t)=N^2_{\phi}(|\alpha_1(t)\rangle|\alpha_2(t)\rangle\langle\alpha_1(t)|\langle\alpha_2(t)|
+|-\alpha_1(t)\rangle|-\alpha_2(t)\rangle\langle-\alpha_1(t)|\langle-\alpha_2(t)|
$$
$$
+\beta_{12}|\alpha_1(t)\rangle|\alpha_2(t)\rangle\langle-\alpha_1(t)|\langle-\alpha_2(t)|
+\beta_{12}^{\ast}|-\alpha_1(t)\rangle|-\alpha_2(t)\rangle\langle\alpha_1(t)|\langle\alpha_2(t)|),
\eqno{(8)}
$$
where $\beta_{12}=\exp[-i\phi-2d^2(|\alpha_1|^2+|\alpha_2|^2)]$,
$d=\sqrt{1-e^{-\gamma{t}}}$, and
$|\pm\alpha_i(t)\rangle=|\pm\alpha_ie^{-\frac{\gamma}{2}t}\rangle$
($i=1,2$). In order to calculate the entanglement of formation of
$\rho(t)$, we introduce two orthogonal vector $|1\rangle_i$ and
$|0\rangle_i$ ($i=1,2$) as follows,
$$
|1\rangle_i=|\alpha_i(t)\rangle_i,~~~|0\rangle_i=\frac{1}{\sqrt{1-\eta^2_i}}
(|-\alpha_i(t)\rangle_i-\eta_i|\alpha_i(t)\rangle_i),
\eqno{(9)}
$$
where $\eta_i=\exp(-2e^{-\gamma{t}}|\alpha_i|^2)$. By making use
of the orthogonal vectors $|1\rangle_i$ and $|0\rangle_i$, we
reexpress $|\alpha_i(t)\rangle_i$ and $|-\alpha_i(t)\rangle_i$ as,
$$
|\alpha_i(t)\rangle_i=|1\rangle_i,~~~|-\alpha_i(t)\rangle_i=\eta_i|1\rangle_i
+\sqrt{1-\eta^2_i}|0\rangle_i.
\eqno{(10)}
$$
Being encoded as above, $\rho(t)$ in Eq.(8) becomes a mixed
two-qubit state,
$$
\rho(t)=N^2_{\phi}\{\frac{1}{2}[(1+(\beta_{12}+\beta_{12}^{\ast})\eta_1\eta_2+\eta^2_1\eta^2_2)
|11\rangle\langle11|+\eta^2_1(1-\eta^2_2)|10\rangle\langle10|
$$
$$
~~~~+\eta^2_2(1-\eta^2_1)|01\rangle\langle01|+(1-\eta^2_1)(1-\eta^2_2)|00\rangle\langle00|]
$$
$$
~~~~+(\beta_{12}+\eta_1\eta_2)\eta_1\sqrt{1-\eta^2_2}|11\rangle\langle10|
+(\beta_{12}+\eta_1\eta_2)\eta_2\sqrt{1-\eta^2_1}|11\rangle\langle01|
$$
$$
~~~~+(\beta_{12}+\eta_1\eta_2)\sqrt{(1-\eta^2_1)(1-\eta^2_2)}|11\rangle\langle00|
+\eta_1\eta_2\sqrt{(1-\eta^2_1)(1-\eta^2_2)}|10\rangle\langle01|
$$
$$
~~~~+\eta_1(1-\eta^2_2)\sqrt{1-\eta^2_1}|10\rangle\langle00|+\eta_2(1-\eta^2_1)
\sqrt{1-\eta^2_2}|01\rangle\langle00|\}+h.c..
\eqno{(11)}
$$
For a mixed two-qubit state, $E$ is equal to
$h(\frac{1}{2}+\frac{1}{2}\sqrt{1-C^2})$, where $h$ is the binary
entropy function $h(x)=-x\log_2x-(1-x)\log_2(1-x)$ and $C$ is the
concurrence defined by[10]
$$
C=\max\{0,\lambda_1-\lambda_2-\lambda_3-\lambda_4\},
\eqno{(12)}
$$
where the $\lambda_i$ ($i=1,2,3,4$) are the square roots of the
eigenvalues in decreasing order of magnitude of the "spin-flipped"
density matrix operator
$R=\rho(\sigma^{y}\otimes\sigma^{y})\rho^{\ast}(\sigma^{y}\otimes\sigma^{y})$.
In Fig.3, we plot the entanglement of formation $E$ as a function
of the degree of decay $d=\sqrt{1-e^{-\gamma{t}}}$ and the
relative phase $\phi$ for (a)$|\alpha|=0.5$ and (b)$|\alpha|=0.2$.
Here we have set $|\alpha_1|=|\alpha_2|=|\alpha|$. The numerical
results show that the entanglement of formation is sensitive with
the relative phase $\phi$ when the amplitude $|\alpha|$ is very
small. In Fig.4, the entanglement of formation $E$ is displayed as
a function of $|\alpha_1|$ and $|\alpha_2|$ for
(a)$\phi=\frac{\pi}{2}$ and (b)$\phi=\pi$ with $d=0.5$. It is
shown that the entanglement of entangled coherent states with
small amplitude $|\alpha|$ is more robust against the decoherence
than the one with large amplitude.

\section * {III. THE PROBABILISTIC TELEPORTATION SCHEME WITH RESTRICTED CLASSICAL INFORMATION}
\hspace*{8mm}In this section, we present the scheme of
probabilistic teleportation via entangled coherent states. In this
scheme, the amount of classical information is restricted to only
one bit.\\
\hspace*{8mm}We assume that Alice wants to teleport a coherent
superposition state
$$
|\psi\rangle_a=A_{+}|\alpha^{\prime}\rangle+A_{-}|-\alpha^{\prime}\rangle
\eqno{(13)}
$$
via the pure entangled coherent channel
$|C(\alpha,\alpha,\pi)\rangle$, where $A_{\pm}$ are unknown
parameters. The state $|\psi\rangle_a$ can be represented as
$$
 |\psi\rangle_a=A^{\prime}_{+}|\psi_+\rangle+A^{\prime}_{-}|\psi_-\rangle,
\eqno{(14)}
$$
where $|\psi_{\pm}\rangle$ are usual even and odd coherent states,
defined by
$$
|\psi_{\pm}\rangle=N_{\pm}(|\alpha^{\prime}\rangle\pm|-\alpha^{\prime}\rangle);
~~~N_{\pm}=[2\pm2\exp(-2|\alpha^{\prime}|^2)]^{-1/2}.
\eqno{(15)}
$$
For teleporting the state $|\psi\rangle_a$, Alice and Bob shares
the quantum channel $|C(\alpha,\alpha,\pi)\rangle$. If
$|C(\alpha,\alpha,\pi)\rangle$ is disposed into the vacuum, it
becomes a quantum noise channel, described by
$$
\rho(t)=N^2_{\alpha}\{|\alpha(t)\rangle|\alpha(t)\rangle\langle\alpha(t)|\langle\alpha(t)|
+|-\alpha(t)\rangle|-\alpha(t)\rangle\langle-\alpha(t)|\langle-\alpha(t)|
$$
$$
~~~~~~+\beta|\alpha(t)\rangle|\alpha(t)\rangle\langle-\alpha(t)|\langle-\alpha(t)|
+\beta^{\ast}|-\alpha(t)\rangle|-\alpha(t)\rangle\langle\alpha(t)|\langle\alpha(t)|\},
\eqno{(16)}
$$
where $N_{\alpha}=(2-2e^{-4|\alpha|^2})^{-1/2}$,
$\beta=-\exp(-4d^2|\alpha|^2)$ and
$|\pm\alpha(t)\rangle=|\pm\alpha{e}^{-\frac{\gamma}{2}t}\rangle$.\\
\hspace*{8mm} In Ref.[8], the standard teleportation schemes via
the entangled coherent states have been studied. The amplitude of
the teleported state in the scheme of Ref.[8] is assumed to
decrease with the decoherence if the entangled coherent state is
disposed into vacuum. Here, we proposed a probabilistic
teleportation scheme as follows. For a fixed $|\psi\rangle_a$,
which Alice want to teleport to Bob, Alice performs a Bell state
measurement on the state $|\psi\rangle_a$ and her part of the
quantum channel. If the state projects the whole system into the
states
$|\Phi_{+}\rangle\langle\Phi_{+}|\otimes[\langle\Phi_{+}|(|\psi\rangle_{aa}\langle\psi|
\otimes\rho(t))|\Phi_{+}\rangle]$, where $|\Phi_{+}\rangle$ is the
one of the following and four Bell states defined by
$$
|\Phi_{\pm}\rangle=\frac{\sqrt{2}}{2}(|\psi_{+}\rangle|\psi_{-}\rangle\pm|\psi_{-}\rangle|
\psi_{+}\rangle),
$$
$$
|\Psi_{\pm}\rangle=\frac{\sqrt{2}}{2}(|\psi_{+}\rangle|\psi_{+}\rangle\pm|\psi_{-}\rangle|
\psi_{-}\rangle),
\eqno{(16)}
$$
then the teleportation is successful, which occurs in the
probabilistic manner. The mean probability of successful
teleportation is
$$
P_s=N^2_{-}N^2_{\alpha}[1+\exp(2e^{-\gamma{t}}|\alpha|^2-4|\alpha|^2)]\exp(-|\alpha^{\prime}|^2
-e^{-\gamma{t}}|\alpha|^2)[\cosh(2e^{-\frac{\gamma}{2}t}|\alpha^{\prime}\alpha|)-1]
$$
$$
~~~+N^2_{+}N^2_{\alpha}[1-\exp(2e^{-\gamma{t}}|\alpha|^2-4|\alpha|^2)]\exp(-|\alpha^{\prime}|^2
-e^{-\gamma{t}}|\alpha|^2)[\cosh(2e^{-\frac{\gamma}{2}t}|\alpha^{\prime}\alpha|)+1],
\eqno{(17)}
$$
here, we have assumed that $\alpha^{\prime}$ and $\alpha$ have the
same phase. In this probabilistic teleportation scheme, Bob need
not know the details of the channel. The classical information
sent from Alice to Bob is just only S (Success) or F (Failure),
which determines whether Bob accept the state or not. In the
following, we calculate the mean fidelity $F$ defined as
follows[13],
$$
F=\overline{_a\langle\psi|\rho_b|\psi\rangle_a}
$$
$$
~~~=2\exp(-|\alpha^{\prime}|^2-e^{-\gamma{t}}|\alpha|^2)\{2(1-\beta)\ln[\frac{\nu(1-\beta\eta)}{\mu(1+\beta\eta)}]\frac{\mu\nu}{\nu(1-\beta\eta)-\mu(1+\beta\eta)}
$$
$$
~~~+(1+\beta)(\mu^2+\nu^2)\frac{\nu^2(1-\beta\eta)^2-\mu^2(1+\beta\eta)^2-2\mu\nu(1-\beta^2\eta^2)\ln[\frac{\nu(1-\beta\eta)}{\mu(1+\beta\eta)}]}{[\nu(1-\beta\eta)-\mu(1+\beta\eta)]^3}\},
\eqno{(18)}
$$
where, $\eta=\exp(-2e^{-\gamma{t}}|\alpha|^2)$,
$\mu=N^2_{+}\cosh^2(|e^{-\frac{\gamma{t}}{2}}\alpha^{\prime}\alpha|)$
and
$\nu=N^2_{-}\sinh^2(|e^{-\frac{\gamma{t}}{2}}\alpha^{\prime}\alpha|)$.
In Fig.5, we plot the mean fidelity $F$ as a function of the
degree of decay $d$ and the initial amplitude $|\alpha|$ of the
entangled coherent states with $|\alpha^{\prime}|=1$. It is seen
that the mean fidelity revivals during the decoherence process in
some situations. This means that the decoherence due to
interaction with the vacuum environment can improve the mean
fidelity in some specific situations. In Fig.6., we show that the
decoherence can enhance the mean fidelity above the classical
limit $2/3$ (The plot range is from $2/3$ to $0.68$).

\section * {IV. CONCLUSION}
\hspace*{8mm}In this Letter, we study the entanglement of
entangled coherent states in the vacuum environment by employing
the entanglement of formation. The results indicate that the
entanglement of formation of the entangled coherent states is
sensitive with the relative phase $\phi$ when the amplitude
$|\alpha|$ is very small. Furthermore, we present a probabilistic
teleportation scheme, in which the amount of classical information
sent by Alice is restricted to one bit. In this probabilistic
teleportation scheme, the decoherence due to interaction with the
environment can enhance the mean fidelity in some specific
situations. It is interesting to verify the probabilistic
teleportation scheme experimentally.
\section * {ACNOWLEDGMENTS}
This project was supported by the National Natural Science
Foundation of China (Project NO. 10174066). The project was also
supported by the Zhejiang Provincial Natural Science Foundation of
China.

\newpage
{\Large\bf Figure Caption}
\begin{description}
\item[FIG.1.]The entanglement of formation $E$ as a function of the amplitude $|\alpha_1|$
and $|\alpha_2|$ with $\phi=\pi$.\\
\item[FIG.2.]The entanglement of formation $E$ as a function of the amplitude $|\alpha|$
and the relative phase $\phi$.\\
\item[FIG.3.]The entanglement of formation $E$ as a function of the degree of decay $d$
and the relative phase $\phi$ for (a)$|\alpha|=0.5$ and (b)$|\alpha|=0.2$.\\
\item[FIG.4.]The entanglement of formation $E$ as a function of the amplitude $|\alpha_1|$ and
$|\alpha_2|$ for (a)$\phi=\frac{\pi}{2}$ and (b)$\phi=\pi$ with the degree of decay $d=0.5$.\\
\item[FIG.5.]The mean fidelity $F$ as a function of the degree of decay $d$ and the initial
amplitude $|\alpha|$ with $|\alpha^{\prime}|=1$.\\
\item[FIG.6.]The same as Fig.5..\\
\end{description}

\begin{figure}
\centering
\includegraphics{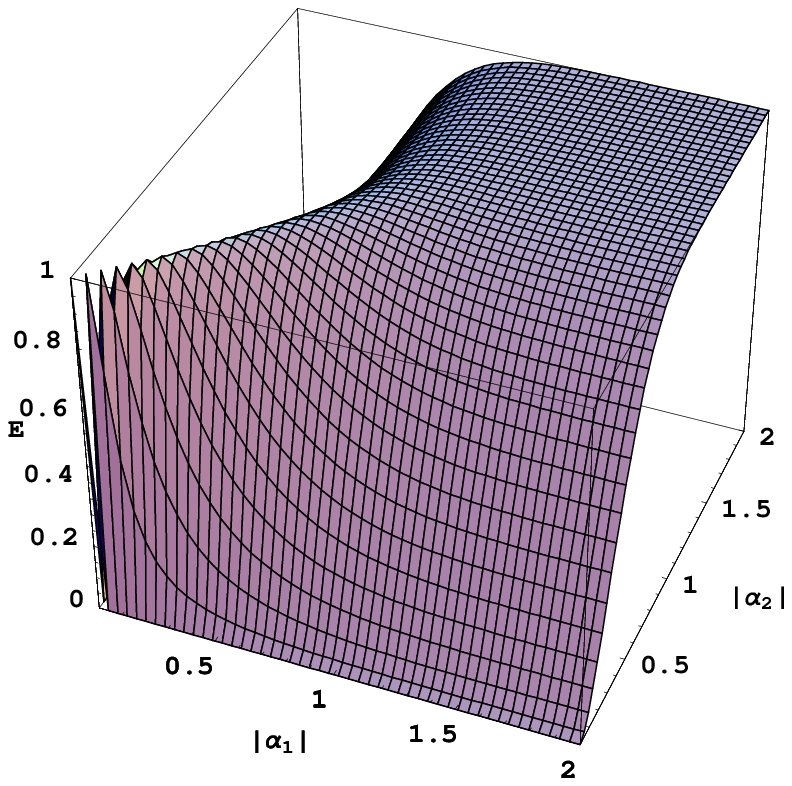}
\caption{The entanglement of formation $E$ as a function of the
amplitude $|\alpha_1|$ and $|\alpha_2|$ with $\phi=\pi$.
\label{Fig1}}
\end{figure}

\begin{figure}
\centering
\includegraphics{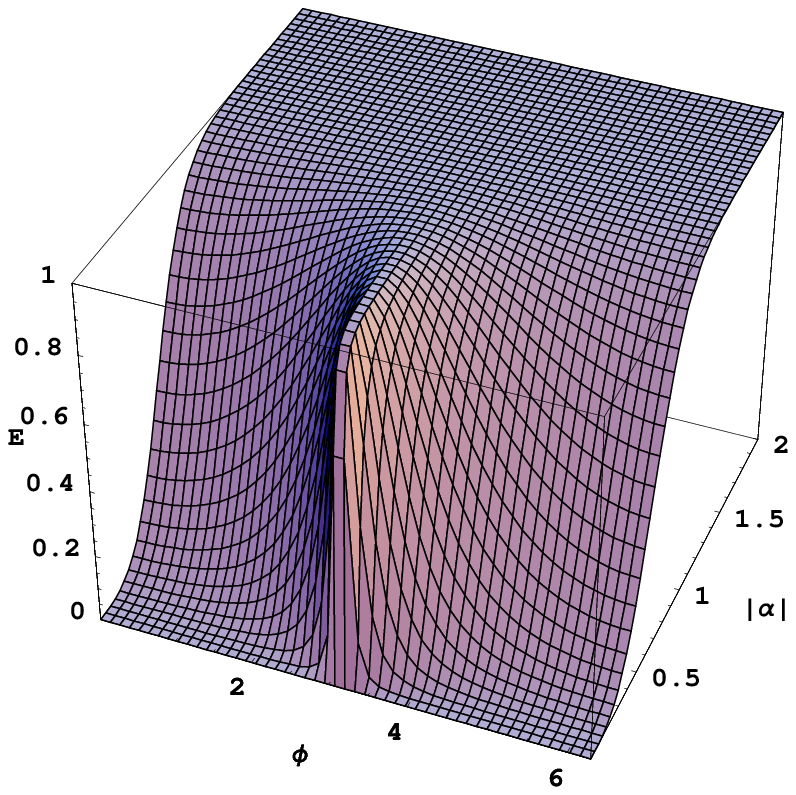}
\caption{The entanglement of formation $E$ as a function of the
amplitude $|\alpha|$ and the relative phase $\phi$. \label{Fig2}}
\end{figure}
\begin{figure}
\centering
\includegraphics{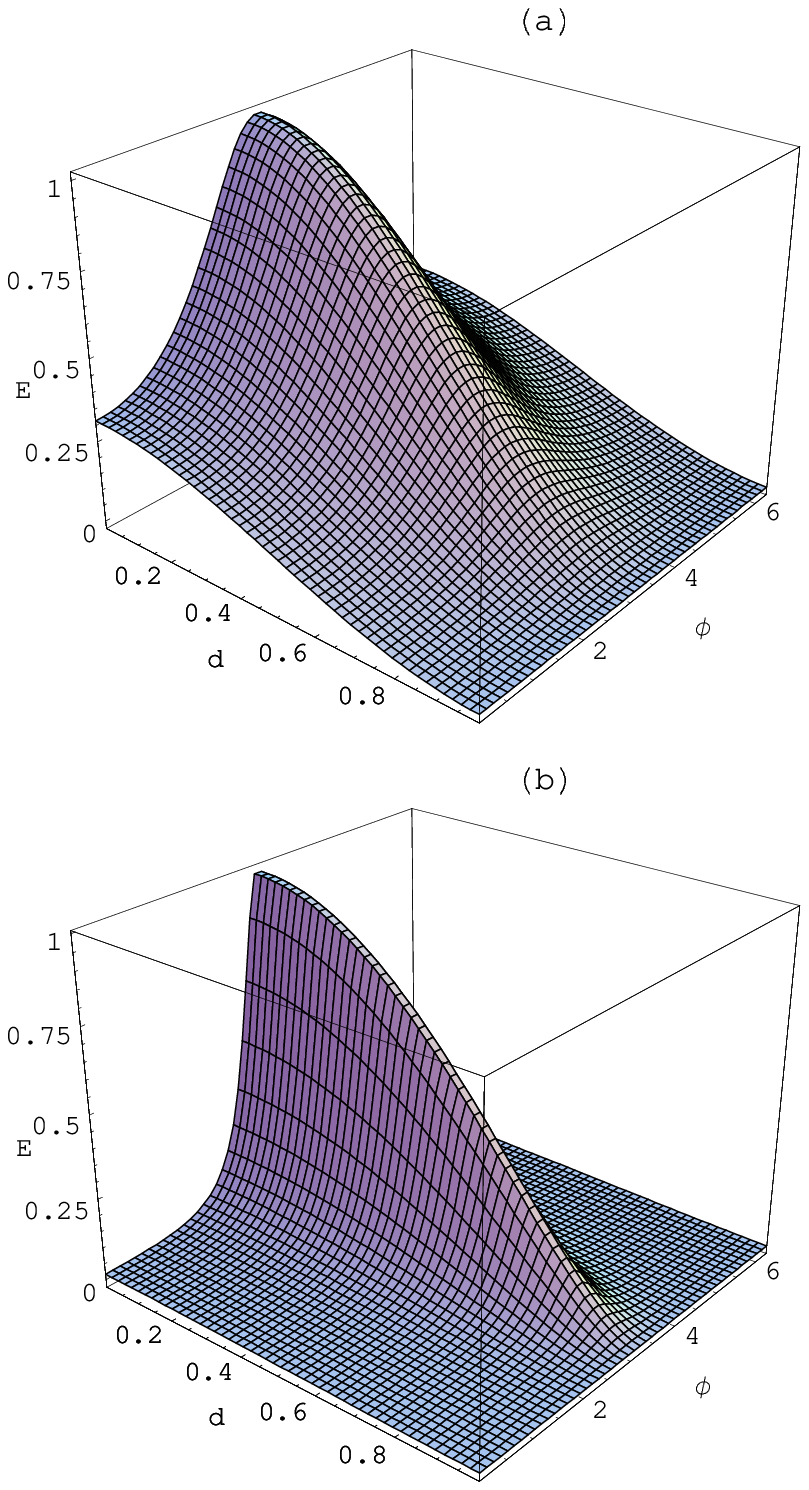}
\caption{The entanglement of formation $E$ as a function of the
degree of decay $d$ and the relative phase $\phi$ for
(a)$|\alpha|=0.5$ and (b)$|\alpha|=0.2$. \label{Fig3}}
\end{figure}
\begin{figure}
\centering
\includegraphics{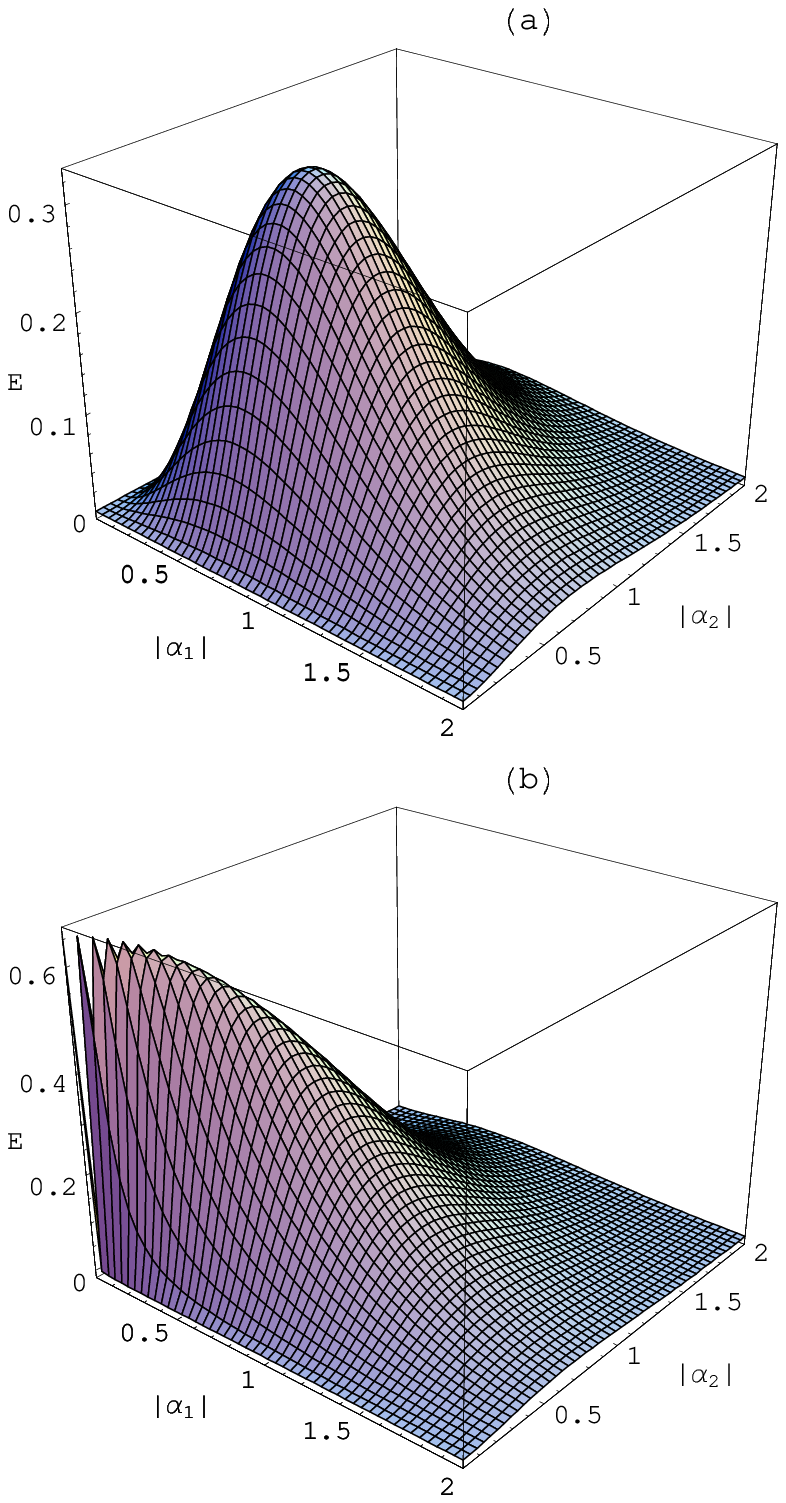}
\caption{The entanglement of formation $E$ as a function of the
amplitude $|\alpha_1|$ and $|\alpha_2|$ for
(a)$\phi=\frac{\pi}{2}$ and (b)$\phi=\pi$ with the degree of decay
$d=0.5$. \label{Fig4}}
\end{figure}
\begin{figure}
\centering
\includegraphics{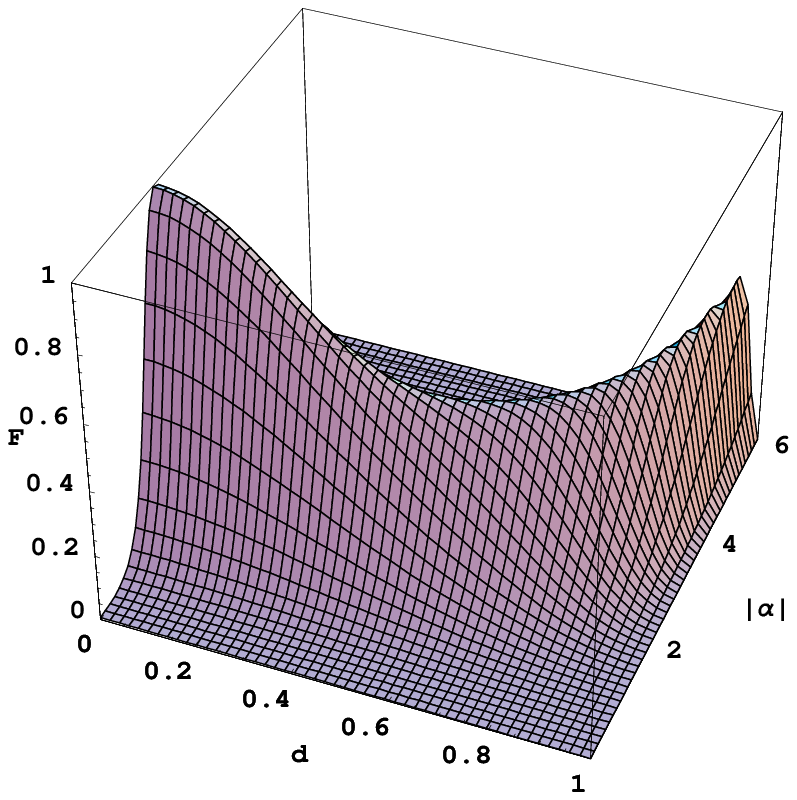}
\caption{The mean fidelity $F$ as a function of the degree of
decay $d$ and the initial amplitude $|\alpha|$ with
$|\alpha^{\prime}|=1$. \label{Fig5}}
\end{figure}
\begin{figure}
\centering
\includegraphics{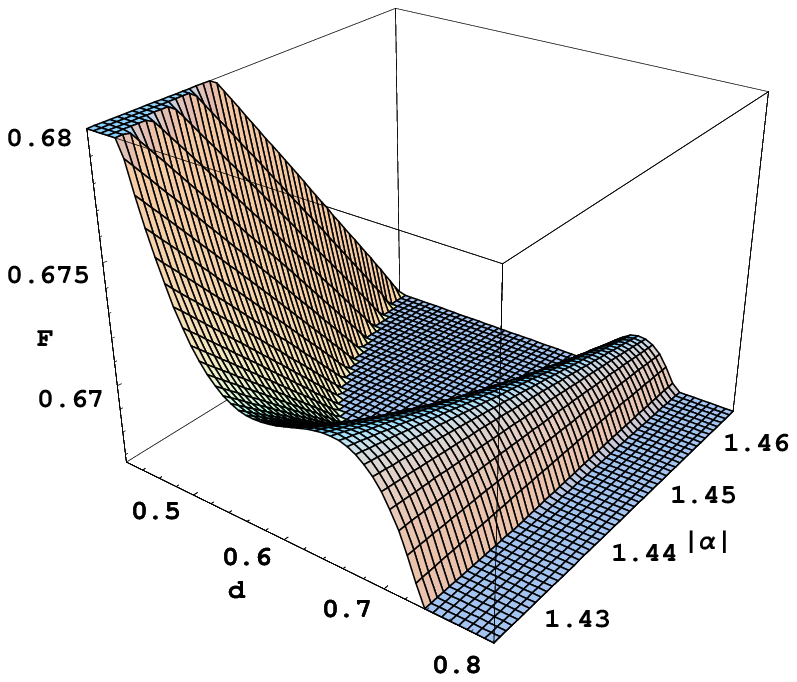}
\caption{The same as Fig.5. \label{Fig6}}
\end{figure}

\end{document}